\documentclass[11pt,a4paper]{article}
\usepackage[hyperref]{emnlp-ijcnlp-2019}
\usepackage{times}
\usepackage{latexsym}

\usepackage{algpseudocode}
\usepackage{algorithm}
\usepackage{amsmath}
\usepackage{amssymb}
\usepackage{hyperref}

\usepackage{url}

\aclfinalcopy 

\title{A Hub-and-Spoke Model for Content-Moderation-at-Scale on an Information-Sharing Platform}

\author{Greg Coppola \\
Think different, again. \\
{\em greg@thinkdifferentagain.art} \\
\\
October 20, 2020
}

\date{}

\begin{document}
\maketitle
\begin{abstract}
One of the most expensive parts of maintaining a modern {\em information-sharing platform} (e.g., {\em web search}, {\em social network}) is the task of {\em content-moderation-at-scale}.
{\em Content moderation} is the {\em binary} task of determining whether or not a given {\em user-created message} meets the {\em editorial team}'s {\em content guidelines} for the site.
The challenge is that the number of messages to check scales with the number of users, which is much larger than the number of moderator-employees working for the given platform.

We show how content moderation can be achieved significantly more cheaply than before, in the special case where all messages are {\em public}, by effectively {\em platformizing} the task of content moderation.
Our approach is to use a {\em hub-and-spoke} model.
The {\em hub} is the core editorial team delegated by the management of the given platform.
The {\em spokes} are the individual users.
The ratings of the editorial team create the {\em labels} for a {\em statistical learning} algorithm, while the ratings of the users are used as {\em features}.

We have implemented a primitive version of this algorithm into our open-source {\em DimensionRank} codebase, found at \mbox{\em thinkdifferentagain.art}.
\end{abstract}

\section{Introduction}
One of the most expensive and labor-intensive parts of running a modern {\em web search} or {\em social network} service is the implementation of {\em content-moderation-at-scale}.
By this, we mean the removal of messages that are deemed in violation of the platform's {\em content guidelines}.
Indeed, if a network has millions or billions of users, and only tens, hundreds or thousands of employees, exhaustive manual annotation of each message is impossible. 
It is empirically much more difficult and costly to scale an employee-based organization, than to scale a social media user-base.

The expense of manual content moderation, to the extent this is possible at all, is a burden on large companies.
But, even worse, it is a prohibition against the formation of new information-platform companies, because this investment in moderation will not be affordable for small companies.

In this paper, we will show a way to {\em radically lessen} the amount of work that needs to be done in order to do content moderation for a network.
We call it the {\em hub-and-spoke} model of moderation.
It crucially leverages {\em statistical methods}, especially {\em deep learning}, as well as the personal neural representations used in the {\em DimensionRank} algorithm and software package \citep{coppola:20}.

\section{Our Context for Content Moderation}
\subsection{Information-Sharing Platform}
In general, we define a {\em platform} as a multi-way channel that connects {\em many producers} to {\em many consumers}, at the same time.
An {\em information-sharing platform} is a platform that users use to share {\em information}.
We refer to a particle of information as a {\em message}.
Each user on an information-sharing platform is can both author messages or receive them, making each user by turns both a producer and a consumer.

\subsection{Content Moderation}
The platform will have a set of {\em content guidelines}, governing whether or not a message is {\em acceptable}, meaning it is allowed on the network, or {\em unacceptable}, meaning it is not allowed on the network.
We assume this is a binary decision, because an information message must either be available on the network, or not.
We will call unacceptable messages {\em toxic}, and use the variable $t$.
So, for a given message, either $t=1$, meaning that the message is toxic, unacceptable and must be removed, or else $t=0$, meaning the message is acceptable.

\subsection{All Messages Public}
We assume that each message is broadcast publicly to the entire network.
Thus, privacy is assumed to be not an issue.
Also, the more times that a message is exposed to users, the more chances users will have to mark it as toxic.

\subsection{User Participation}
We assume that {\em some} users participate in the moderation scheme, by being willing to mark messages as either $t=1$ or $t=0$.
We do not require {\em all} users to participate in moderation, but this algorithm {\em does require} crucially that, for each message, at least one reliable user, and preferably multiple users, will be willing to mark a message as toxic if they see it.
In practice, on most existing platforms, we believe that under-reporting has not been a problem, and there has instead been the problem of {\em too many} people rating things toxic, rather than too few.

\subsection{Neural Networks for an Inherently Fuzzy Task}
A content moderation policy may in some sense attempt to give ``clear rules'' for what is a violation or not.
However, policies that reference the {\em linguistic} or {\em visual} human cognitive systems must inevitably refer to topics that require a judgment decision that cannot be written in a rule-based way.
This is because rule-based {\em natural language processing} and rule-based {\em vison-processing} both do not work.
Both fields require some amount of machine learning.
Thus, we believe it appropriate to model task of content-moderation using a {\em statistical process}, e.g., a {\em neural network}, which can operate based on its own statistical representation of the data.

\section{The Hub-and-Spoke Model}
Our approach to radically reduce costs is to use a {\em hub-and-spoke} model.

\subsection{Hubs and Spokes}
The essence of the {\em hub} and {\em spoke} metaphor is that the {\em hub} is a central, organizing node in a network system, and the {\em spokes} play a role by co-ordinating directly with the hub.
This would be as opposed to a completely {\em decentralized} network, in which there is no notion of a ``central'' node.
In our case, we need a central node, because this corresponds to the editorial team, whose need to supervise the network is the initial motivation for this work.

So, the hub corresponds to the editorial team, and the spokes to the users.
As we said, the traditional problem plaguing content moderation is that the number of messages scales with the number of users, which is much larger than the editorial team.
But, we turn this crisis into opportunity, by {\em leveraging} the activity and intelligence of users.
The users and are invited to flag content as {\em toxic}.
But, we are not so rash as to ban a message because some individual user calls it toxic.
We instead leverage {\em algorithms} and {\em statistical learning} to predict what the editorial team {\em would think}, if they had rated it, given its independent characteristics (text and image), and given the certain set of users have rated either $t=1$ or $t=0$.

{\em Conclusion}
We only take a message down if either the editorial team literally marks something as toxic, or if the algorithms believe they would do so, with sufficiently high probability.

\subsection{Labeled Examples}
The goal is to label an individual broadcast message $m$, as either toxic ($t=1$), or acceptable ($t=0$).
There are two ways we can get a label.
First, a user (spoke) can label a message.
Second, the editorial team (hub) can label a message.
The {\em gold label} will come from the editorial team.
The labels from the users will be used as {\em features}.

\subsection{Feature Sets}
The primary input features are (anonymized) identities of the users who have created labels, either $t=1$ or $t=0$.
Because we are using the {\em DimensionRank} algorithm, we will be able to represent each user as a {\em representation vector} in a multi-dimensional space.
However, the currently implemented partial version of the algorithm simply uses the {\em number} of users as a signal to flag manual review.
Aside from using social features, we can also represent a message using a neural representation of the text or image itself.

\subsection{Training and Prediction}
A new {\em training example} is created each time that the editorial team makes a new label.
A new {\em prediction example} is created each time that a user makes a new label $t=1$.
If a message truly is toxic, we expect it to get multiple votes $t=1$, and we will create a new prediction example each time.
The more votes for toxic a message receives, the more likely we would expect it to be found toxic.

\section{Implementation}
We have implemented a primitive version of this algorithm into our open-source {\em DimensionRank} codebase, which can be found at \mbox{\em thinkdifferentagain.art}.
Currently, we use the number of times each article has been flagged to prioritize manual review.
However, we plan to implement more sophisticated versions, as the number of users grows.
We will provide an update on metrics when they are available.

\bibliography{emnlp-ijcnlp-2019}
\bibliographystyle{acl_natbib}

\appendix

\end{document}